\documentclass[a4paper,11pt]{article}
\usepackage{jheppub} 
\usepackage{lineno}
\usepackage{amsmath}
\usepackage{slashed}
\usepackage[svgnames]{xcolor}

\newcommand{\nb}{\bar{n}}
\newcommand{\Qb}{\bar{Q}}
\newcommand{\Amp}{\mathcal{M}}

\usepackage[normalem]{ulem}
\usepackage{marginnote}

\preprint{MIT-CTP 5872}

\title{\boldmath Relativistic corrections to exclusive photoproduction of Quarkonia near-threshold}

\author[1]{Sarah K. Blask,}
\author[1]{Sean Fleming,}
\affiliation[1]{Department of Physics, University of Arizona, Tucson, Arizona 85721, USA}
\author[2]{Thomas Mehen,}
\author[2]{Jyotirmoy Roy,}
\affiliation[2]{Department of Physics, Duke University, Durham, North Carolina 27708, USA}
\author[3,4]{Iain~W.~Stewart,}
\author[3]{Fanyi Zhao}
\affiliation[3]{Center for Theoretical Physics -- a Leinweber Institute, Massachusetts Institute of Technology, Cambridge, MA 02139, USA}
\affiliation[4]{University of Vienna, Faculty of Physics, Boltzmanngasse 5, A-1090 Wien, Austria}

\emailAdd{sarahkblask@arizona.edu, spf@arizona.edu, 
jyotirmoy.roy@duke.edu, iains@mit.edu
}

\abstract{Non-relativistic QCD (NRQCD) is used to calculate the relativistic correction to the amplitude for exclusive photoproduction of vector Quarkonia in the near-threshold region within the generalized parton distribution (GPD) framework. The relativistic corrections are found to be large for $J/\psi$, and lead to a breakdown of the GPD moment expansion near threshold. Cross-sections for both $J/\psi$ and $\Upsilon$ are calculated with the former being compared to the data. We also demonstrate the presence of endpoint divergences for the relativistic correction away from the near-threshold regime.}

\makeatletter
\gdef\@fpheader{}
\makeatother

\begin{document}
\maketitle
\flushbottom

\section{Introduction}
An important goal of quantum chromodynamics (QCD) is to explain the properties of nucleons in terms of the underlying partons. One step in this direction is elucidating the spatial distribution of the partons in a nucleon encoded in non-perturbative quantities called generalized parton distributions (GPDs) \cite{Muller:1994ses,Radyushkin:1997ki,Ji:1996ek}. These GPDs can be accessed in exclusive reactions like deeply virtual Compton scattering and deeply virtual meson production. With the already running experiments at Thomas Jefferson Laboratory (JLab) \cite{Adderley:2024czm} and the upcoming Electron-Ion Collider (EIC) \cite{AbdulKhalek:2021gbh}, significant progress will be made in precision measurement of these quantities. Thus higher order and higher power calculations are warranted on the theoretical side. In the present work, we take a first step towards a full analysis of power corrections by determining a unique subset of such corrections for processes with a heavy quarkonia.  

Exclusive production of a vector meson has been studied in different kinematical limits using various methods. Within the perturbative QCD (pQCD) framework, calculations have been performed for the high energy and/or small-$x$ limit in the leading logarithmic approximation within the color dipole \cite{Nikolaev:1990ja,Mueller:1993rr} and $k_T$-factorization approaches \cite{Ryskin:1992ui,Brodsky:1994kf,Kuraev:1998ht,Ivanov:1999pb,Ivanov:2002ef}. For heavy vector mesons (HVM), exclusive electroproduction and photoproduction have also been studied at next-to-leading order (NLO) \cite{Ivanov:2004vd,Chen:2019uit,Flett:2021ghh,Flett:2024htj} based on the collinear factorization method \cite{Collins:1996fb} when $W^2\gg M_V^2,-t$. Here $W$ is the invariant mass of the photon-proton system, $M_V$ is the mass of the HVM, and $t$ is the invariant momentum transfer squared. For HVM production in the near-threshold limit, we also implicitly assume the hierarchy $M_V \gg M_N$, where $M_N$ is the nucleon mass.
The other kinematic region is the near-threshold region, characterized by $W^2 \sim M_V^2 \gg -t$.  In both of these limits the amplitude has been shown to factor into gluon GPDs for both electroproduction \cite{Boussarie:2020vmu} and photoproduction \cite{Guo:2021ibg,Sun:2021pyw,Guo:2023pqw} at leading power (LP). Moreover in \cite{Boussarie:2020vmu,Guo:2021ibg} it has been shown that the amplitude in the threshold limit is directly related to gluon gravitational form factors (GFFs). 

In this paper we study the relativistic correction to photoproduction of vector Quarkonia near-threshold following the strategy of \cite{Guo:2023pqw}, where the nonperturbative dynamics of the proton is encoded in the moments of GPD. Similar relativistic corrections have been studied before, but for electroproduction in the high-energy/small-$x$ limit \cite{Hoodbhoy:1996zg}, which we will not study here. 
As we will show, relativistic corrections are large in the near-threshold limit for photoproduction and thus need to be taken into account to explain data from the GlueX collaboration \cite{GlueX:2019mkq,GlueX:2023pev} in the low $W$ region.

The rest of the paper is organized as follows. In Sec.~\ref{sec:amplitude} we discuss the collinear factorization and Non-relativistic QCD framework and use it to calculate the relativistic correction to the amplitude. In Sec.~\ref{sec:Cross-section} we show the cross-section for both $J/\psi$ and $\Upsilon$, where we compare the former with experimental results. Finally in Sec.~\ref{sec:Summary} we provide a summary and outlook for future work. 

\section{Amplitude for $\gamma(q) + N(p) \rightarrow V(P_V) + N(p^\prime)$}
\label{sec:amplitude}

Exclusive photoproduction of quarkonium is mediated via two-parton exchange and, unlike deep inelastic scattering, the incoming nucleon does not break apart but instead recoils with a different momentum. The production of the $Q\Qb$ pair and its subsequent recombination into an HVM state can be calculated in the nonrelativistic QCD (NRQCD) framework \cite{Caswell:1985ui,Bodwin:1994jh,Braaten:1993rw,Fleming:1997fq,Luke:1999kz,Brambilla:1999xf,Bodwin:2007ga,Copeland:2023wbu} which is an expansion in $\alpha_s$ and $v$, the relative velocity of the quark-antiquark pair. Projecting the $Q\Qb$ pair onto the HVM gives the NRQCD long distance matrix elements (LDMEs). At leading order in the $v$ expansion there is only one LDME which is related to the HVM wavefunction at the origin. Thus using NRQCD and the collinear expansion, the amplitude for this process can be expressed in a factored form consisting of a short-distance hard matching coefficient, the LDME, and the gluon GPD which parametrizes the dynamics of gluons in the recoiling proton at the nonperturbative scale.

\subsection{Kinematics} 
For the purpose of this calculation, we define the lightcone coordinates in terms of the collinear vector $n^{\mu}=(1,0,0,1)$ and the anticollinear vector $\nb^{\mu}=(1,0,0,-1)$. The calculation is most conveniently done in the Ji frame, where the mean of the large lightcone momentum component of the incoming and outgoing proton is taken along $n^\mu$. In the near-threshold limit, the momentum decomposition for the incoming nucleon $p$, the photon $q$, the outgoing nucleon $p^\prime$ and the HVM $P_V$ in terms of the light-cone basis vectors is:
\begin{align}
\label{eq:mom_decomposition}
    & p^{\mu}=(1+\xi) P^{-}\frac{n^{\mu}}{2} +(1-\xi) \frac{M_N^2-\frac{t}{4}}{P^-} \frac{\nb^\mu}{2} -\frac{\Delta^{\mu}_{\perp}}{2} 
   \,, \nonumber \\
    & q^{\mu}=-\frac{2\xi P^-}{M^2_V} \Big[ (1-\xi^2) \frac{t}{4}+ \xi^2 M_N^2 \Big]\frac{n^{\mu}}{2}+\frac{M^2_V}{2\xi P^{-}} \Big[1+ 4\xi^2 \frac{M_N^2-\frac{t}{4}}{M^2_V} \Big] \frac{\nb^{\mu}}{2} +\frac{\Delta^{\mu}_{\perp}}{2}
    \,,\nonumber \\
    & p^{\prime \mu}=(1-\xi) P^- \frac{n^{\mu}}{2} +(1+\xi) \frac{M_N^2-\frac{t}{4}}{P^-} \frac{\nb^\mu}{2}+\frac{\Delta^{\mu}_{\perp}}{2}
    \,,\nonumber \\
    & P_{V}^{\mu}=\frac{2\xi P^-}{M^2_V} \Big[M^2_V -(1-\xi^2) \frac{t}{4} -\xi^2 M_N^2\Big] \frac{n^{\mu}}{2}+\frac{M^2_V}{2\xi P^-}\frac{\nb^{\mu}}{2} -\frac{\Delta^{\mu}_{\perp}}{2}
   \,,
 \end{align}
where $p^2=p^{\prime\,2}=M_N^2$, $q^2=0$, and $P_V^2=M_V^2$.
We take the independent kinematic variables to be $P^-$, $\xi$ and $t$.
Here $P^-=\nb \cdot (p^{\prime}+p)/2$, is the average of the largest momentum components of the incoming and outgoing nucleons, $t=(p^\prime -p)^2$ is the square of the momentum transfer, and $\xi=-\frac{\nb \cdot (p^{\prime}-p)}{\nb \cdot (p^{\prime}+p)}$, the skewness parameter, parametrizes the change in the collinear momentum of the nucleon. 
Note that the magnitude of the perpendicular component is constrained to be $\Delta_\perp^2=(1-\xi^2) t+4\xi^2 M_N^2$. 
To obtain the decomposition in Eq.~(\ref{eq:mom_decomposition}) we have taken $-t, M_N^2 \ll M_V^2$ and only retained terms up to $O\left( M_N^2/M^2_V\right)$ and $O\left( t/M^2_V\right)$.
To this order, the invariant mass $W^2 \equiv (q+p)^2 = M_V^2 (1+\xi)/(2\xi) + M_N^2 (1+2\xi)- t(1+\xi)/2$.

The threshold region can be defined by the limit where $1-\xi \sim M_N/M_V \sim t/M_V^2 \ll 1$.
In this region~\cite{Guo:2021ibg}, $\xi\to 1$, $W^2\simeq (M_V+M_N)^2$ and $-t \sim O (M_N M_V)$, where this last scaling relation can be seen most easily from the kinematics bounds $-t_{-} <-t< -t_{+}$ where 
\begin{align}
\label{eq:trange}
-t_{\pm} =& \frac{(W^2 + M_N^2) (W^2 - M_V^2 + M_N^2)}{
 2 W^2} \\
 & \pm \frac{(W^2 - M_N^2)\sqrt{(W^2 - (M_V + M_N)^2) (W^2 - (M_V - M_N)^2)}}{2 W^2} -2M_N^2 \,.
 \nonumber
\end{align}
These bounds are Lorentz invariant but are derived most easily in the center-of-mass frame.  This has implications for $J/\psi$ since kinematic corrections (dependence of the hard matching coefficient on $t$), and subleading GPD corrections (which go as $-t/W^2 \sim O (M_N/M_V) $), are the same order as the $v^2$ relativistic corrections. For the present work we do not consider these power corrections and so terms which are proportional to $t$ and $M_N^2$ in Eq.(\ref{eq:mom_decomposition}) are neglected in the calculation of the hard matching coefficient.

\subsection{Projection onto Nonperturbative Matrix Elements}
\label{Sec:NonperturbativeME}

At lowest order in $\alpha_s$, the on-shell heavy quark-antiquark pair is produced via two gluon exchange in the partonic processes, which correspond to $\gamma(q) + g(p_1) \rightarrow g(p_2) +Q(p_3)+\Qb(p_4)$ in the DGLAP region ($|x| > |\xi|$) or
$\gamma(q) + g(p_1) + g(-p_2) \rightarrow  Q(p_3)+\Qb(p_4)$ in the ERBL region ($|x| < |\xi|$).  
The matrix element which extracts gluons from the proton is given by the gluon GPD correlator $F^g(x,\xi,t)$, defined by
\begin{equation}
\label{GPDoperator}
    F^g(x,\xi,t)=\frac{1}{(P^-)^2}\int \frac{d\lambda}{2\pi} e^{i \lambda x} \, \bar{n}_{\mu} \bar{n}_{\nu} \left\langle p' \left| G^{\mu \sigma}_{a} \left( -\frac{\lambda \bar{n}}{2} \right) 
   W\Bigl[-\frac{\lambda\bar{n}}{2}, \frac{\lambda\bar{n}}{2} \Bigr]
   G^{\nu}_{a\sigma} \left(\frac{\lambda \bar{n}}{2} \right) \right| p \right\rangle,
\end{equation}
where $W[b,a]=P\exp( ig \int_a^b ds \cdot A )$ denotes a straight Wilson line from $a$ to $b$.
To probe this GPD operator it suffices to take the two gluons to have momenta
\begin{equation}
    p_1^{\mu}=(x+\xi)P^- \frac{n^{\mu}}{2}, \qquad p_2^{\mu}=(x-\xi)P^- \frac{n^{\mu}}{2}
  \,.
\end{equation}
In the GPD factorization theorem the momentum fraction $x$ will appear in both the hard matching coefficient and the GPDs, and is integrated over.
The momenta of the heavy quark-antiquark pair can be expressed in terms of the total momentum $P_V $ and the relative momentum of the pair $k\sim m_Q v$, with $m_Q$ the heavy quark mass:
\begin{align}
    & p_3=\frac{1}{2}P_V+k \nonumber \\
    & p_4=\frac{1}{2}P_V-k.
\end{align}

\begin{figure}[htb!]
    \centering
    \includegraphics[trim={2cm 8cm 0 8cm},clip, scale=0.85]{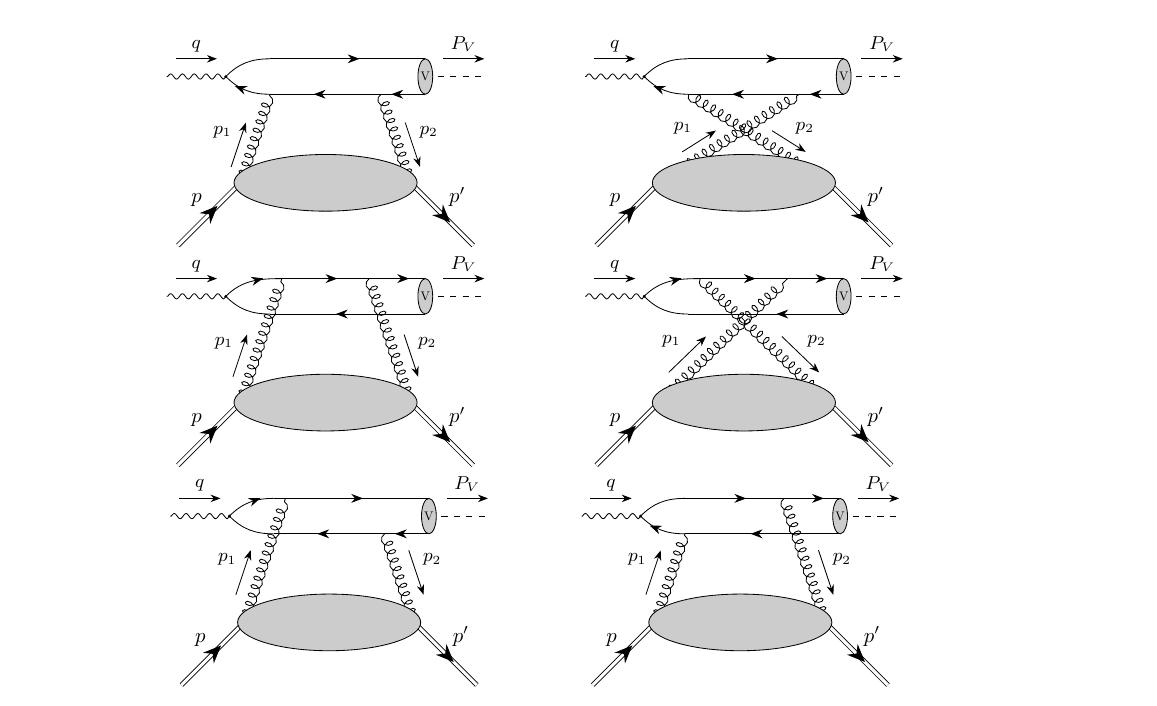}
    \caption{Feynman diagrams required for the calculation of the cross section both at leading power and with relativistic corrections.}
    \label{fig:Feynman_Diagrams}
\end{figure}

There are six Feynman diagrams contributing to the amplitude as shown in Fig.\ref{fig:Feynman_Diagrams}. 
The amplitude can in general be written as
\begin{align}
\label{eq:amplitude}
    \mathcal{M} &=\bar{u}(p_3) \mathcal{A}^{(\alpha \beta)} v(p_4) \varepsilon_\alpha(p_1) \varepsilon^{*}_\beta(p_2) \nonumber \\
    &=\text{Tr}\left[ \mathcal{A}^{(\alpha \beta)} v(p_4) \bar{u}(p_3)\right] \varepsilon_\alpha(p_1) \varepsilon^{*}_\beta(p_2) \nonumber \\
    &\equiv \mathcal{M}^{(\alpha \beta)} \varepsilon_\alpha(p_1) \varepsilon^{*}_\beta(p_2) \,,
\end{align}
where $\mathcal{A}$ is a matrix that acts on spinors with both Dirac and color indices and $\varepsilon_\alpha(p_1),\varepsilon^{*}_\beta(p_2)$ are the polarization vectors for the incoming and outgoing gluon respectively. For later convenience we have indicated the Lorentz indices contracting with the gluon polarization vectors in parentheses. By replacing $v(p_4) \bar{u}(p_3)$ with projection operators one can project the $Q\Qb$ state onto a particular spin and color configuration. For the spin triplet, color singlet configuration the projection operator is
\begin{equation}
    \Pi^{Q\Qb}_{3}(p_4, p_3)=\frac{1}{\sqrt{2}M_V (M_V+2 m_Q)}(\slashed{p}_{4}-m_Q)\slashed{\varepsilon}^{*}_{V} (\slashed{P}_{V}+M_V)(\slashed{p}_3+m_Q) \otimes \left(\frac{1}{\sqrt{N_c}} \bf{1} \right)
  ,
\end{equation}
where $N_c=3$, $\varepsilon^{*}_{V}$ is the quarkonium spin polarization vector satisfying $\varepsilon^{*}_{V} \cdot \varepsilon_{V}=-1$ and $P_{V} \cdot \varepsilon_V=0$, and $\bf{1}$ in the last factor is the unit color matrix. We also express $M_V$ in terms of the heavy quarkmass $m_Q$ via the relation $M_V^2=4(m_Q^2-k^2)$. Using the projection operator the matrix element takes the form
\begin{align}
\label{eq:amp_expansion}
    \mathcal{M}^{(\alpha \beta)}[Q\Qb(S=1)] =& \text{Tr} \left[ \mathcal{A}^{(\alpha \beta)} \Pi^{Q\Qb}_{3} \right] = \mathcal{M}^{(\alpha \beta)}_{\mu \nu} \varepsilon_{\gamma}^{\mu} \varepsilon_{V}^{*\nu} 
 \nonumber \\
    =& \left( \mathcal{M}^{(0)(\alpha \beta)}_{\mu \nu}+\mathcal{M}^{(1)(\alpha \beta \rho)}_{\mu \nu} k_{\rho}+\mathcal{M}^{(2)(\alpha \beta)}_{\mu \nu \rho \sigma }k^{\rho}k^{\sigma}+\dots \right) \varepsilon_{\gamma}^{\mu} \varepsilon_{V}^{*\nu}
\end{align}
where in the second line the amputated amplitude $\mathcal{M}^{(\alpha \beta)}_{\mu \nu}$ has been expanded in powers of $k$. The amplitude for the spin-triplet quarkonium state $V$ at leading order in $v$ is 
\begin{equation}
\mathcal{M}^{(\alpha \beta)}[V]=\left( \frac{\langle \mathcal{O}_{1}(^3S_1) \rangle_{V}}{N_c M_V } \right)^{1/2} \mathcal{M}^{(0)(\alpha \beta)}_{\mu \nu} \varepsilon_{\gamma}^{\mu} \varepsilon_{V}^{*\nu}\,,
\end{equation}
where $\langle \mathcal{O}_{1}(^3S_1) \rangle_{V} \equiv \langle V | \chi^\dagger \boldsymbol{\sigma} \psi \cdot \chi^\dagger \boldsymbol{\sigma} \psi | V \rangle $ is the nonrelativistic LDME for the color singlet, spin triplet operator. 
Matching the amplitude onto the gluon GPD correlator $F^g(x,\xi,t)$
is achieved by the replacement 
\begin{equation}
\label{eq:GPD_projector}
    \epsilon_a^{\alpha}(p_1) \epsilon_b^{*\beta}(p_2) \rightarrow -\frac{\delta_{ab}}{(N_c^2-1)}\frac{g^{\alpha \beta}_{\perp}}{2(d-2)} \frac{F^g(x,\xi,t)}{(x-\xi+i\epsilon)(x+\xi-i\epsilon)}\,,
\end{equation}
where $a, b$ are color indices in the adjoint representation, $g^{\alpha \beta}_{\perp} \equiv g^{\alpha \beta}-\frac{1}{2}n^{\alpha}\nb^{\beta}-\frac{1}{2}\nb^{\alpha}n^{\beta}$ and the factor of $d-2$ in the denominator is to average over the transverse gluon polarizations in $d$-dimensions.

\subsection{Relativistic Correction and Gravitational Form Factors}

Before we discuss the relativistic corrections it is instructive to review the leading power calculations. There are six diagrams which contribute to the amplitude at leading order in $\alpha_s$. Using the projections discussed in Sec.\ref{Sec:NonperturbativeME}, the amplitude can be readily evaluated
\begin{align}
    \Amp^{(0)}[V] = \frac{8\sqrt{2}\pi g_{e} e_{Q} \alpha_s}{3}\left( \frac{\langle \mathcal{O}_{1}(^3S_1) \rangle_V}{M_V} \right)^{1/2} \left( \varepsilon_\gamma^\perp \cdot \varepsilon_V^{*\perp} \right) \int_{-1}^{1} dx \, C^{(0)}(x,\xi) F^{g}(x,\xi,t)
\end{align}
where the hard matching coefficient is
\begin{equation} \label{eq:C0}
    C^{(0)}(x,\xi)=-\frac{1}{2m_Q}\frac{1}{(x-\xi+i\epsilon)(x+\xi-i\epsilon)}.
\end{equation}
This reproduces the LO amplitude in \cite{Guo:2021ibg} with the identification of the NRQCD matrix element with the wavefunction at the origin $\langle \mathcal{O}_{1}(^3S_1) \rangle_V =2 N_c |\psi(0)|^2$ and $M_V=2m_Q$ at leading power in the vacuum saturation approximation.
Parametrizing the gluon GPD correlator in terms of a nucleon spinor $U(p)$ we can write
\begin{equation}
\label{eq:gluonGPD}
    F_g(x,\xi,t)=\frac{1}{P^-}\left[ H_g(x,\xi,t) \bar{U}(p^\prime) \frac{\slashed{\nb}}{2} U(p)+E_g(x,\xi,t)\bar{U}(p^\prime) \frac{i \sigma^{\mu \nu} (p^\prime -p)_{\nu}}{2M_N} U(p) \frac{\bar{n}_\mu}{2}\right] 
  ,
\end{equation}
where $H_g$ and $E_g$ are (scalar) GPDs. 

 Since in the near-threshold limit $\xi \sim 1$ one can expand the hard matching coefficient about $|x|\ll |\xi|$, 
and express the convolution in terms of even moments of the GPD
\begin{align}
    \Amp^{(0)}[V] = \frac{8\sqrt{2}\pi g_{e} e_{Q} \alpha_s}{3}\left( \frac{\langle \mathcal{O}_{1}(^3S_1) \rangle_V}{M_V} \right)^{1/2} \left( \varepsilon_\gamma^\perp \cdot \varepsilon_V^{*\perp} \right) G^{(0)}_{LP}(t,\xi)\,,
\end{align}
where $G^{(0)}$ denotes the leading power projection of the GPD correlator, which itself has a Taylor series in $x^2$ moments 
\begin{equation}
\label{eq:int_gpd}
    G^{(0)}(t,\xi) \equiv \frac{1}{2m_Q\xi^2}\int_{-1}^{1} dx \, F_g (x,\xi,t)+\dots
   \,.
\end{equation}
This zero'th moment can then be expressed in terms of the moments of $H_g$ and $E_g$
\begin{align} \label{eq:zeromoment}
    & \int_{0}^{1} dx \, H_g(x,\xi,t) =A_g(t)+(2\xi)^2 C_{g}(t) \equiv H_2(t,\xi) 
  \,, \nonumber \\
    & \int_{0}^{1} dx \, E_g(x,\xi,t) =B_g(t)-(2\xi)^2 C_{g}(t) \equiv E_2(t,\xi)\,.
\end{align}
Here the terms $A_g, B_g$ and $C_g$ are the gravitational form factors relating the gluon stress-energy tensor to the GPDs.

In the above equations we have retained terms only up to the first term in the moment expansion. 
Both the approximation of expansion in $x$ and retaining only the zeroth moment in the near-threshold region is justified in \cite{Hatta:2021can} based on the asymptotic form of the GPD correlator  
\begin{equation} \label{eq:Fasym}
    F_g^{\text{asym}}(x,\xi,t) \propto \left( 1-\frac{x^2}{\xi^2} \right)^2\theta \left( 1-\frac{x}{\xi} \right)
 \,.
\end{equation}
This applies equally well for $H_g$ and $E_g$, just with different proportionality constants.
In the threshold limit, we can set the scale $\mu\sim M_V$, which is in the ultraviolet and much larger than the other kinematic scales, and hence this asymptotic form is a reasonable approximation. 
 By neglecting higher order $x$ moments a contribution of $25\%$ is being neglected as estimated from the asymptotic form of the GPD correlator
\begin{equation} 
\label{eq:onequarter}
    \frac{\sum_{n \ge 1}^{\infty}\int_{-1}^{1} dx \, \frac{x^{2n}}{\xi^{2n}}F_{g}^{\text{asym}}(x,\xi,t)}{\int_{-1}^{1} dx \, F_{g}^{\text{asym}}(x,\xi,t)}=\frac{1}{4}\,.
\end{equation}
Note that the asymptotic form in Eq.~(\ref{eq:Fasym}) is used to estimate the size of these power corrections, but our analysis below does not rely on this approximation. 

%
%

The relativistic correction to the LO amplitude is obtained by projecting the full amplitude onto the higher order LDME \cite{Bodwin:2002cfe,Braaten:2002fi}
\begin{equation}
    \langle \mathcal{P}_{1}(^3S_1) \rangle_{V} = \left\langle V \left| \frac{1}{2} \left[\chi^\dagger \boldsymbol{\sigma} \psi \cdot \psi^\dagger \boldsymbol{\sigma} \left( -\frac{i}{2}\boldsymbol{ \overleftrightarrow{D}} \right)^2\chi +\chi^\dagger \boldsymbol{\sigma} \left( -\frac{i}{2}\boldsymbol{ \overleftrightarrow{D}} \right)^2 \psi \cdot \psi^\dagger \boldsymbol{\sigma} \chi \right] \right| V \right\rangle
 .
\end{equation}
This can be conveniently expressed in a manner proportional to the original matrix element $\langle \mathcal{O}_{1}(^3S_1) \rangle_V$ by defining the ratio
\begin{equation}
    \langle v^2 \rangle_V \equiv \frac{\langle \mathcal{P}_{1} (^3S_1)\rangle_{V}}{m_Q^2 \langle \mathcal{O}_{1}(^3S_1) \rangle_{V}}\,.
\end{equation}
Phenomenologically this ratio can be determined using the Gremm-Kapustin relation \cite{Gremm:1997dq}, which gives
\begin{equation} \label{eq:v2V}
    \langle v^2 \rangle_V \approx \frac{M_V^2-4m_Q^2}{4m^{2}_{Q}}.
\end{equation}
Using these results amplitude with the leading relativistic corrections is given by
\begin{align}
\label{eq:RelCorrection}
    \Amp[V]  &= \frac{8\sqrt{2}\pi g_{e} e_{Q} \alpha_s}{3}  \left( \frac{\langle \mathcal{O}_{1}(^3S_1) \rangle_V}{M_V} \right)^{1/2} \left( \varepsilon_\gamma^\perp \cdot \varepsilon_V^{*\perp} \right) 
    G(\xi,t)\,,
     \nonumber \\  
     G(\xi,t) &\equiv \int_{-1}^{1} dx \left[ C^{(0)}(x,\xi) +C^{(2)} (x,\xi) \right] F^{g}(x,\xi,t)\,,
\end{align}
where our result for the subleading hard function and ratio of NRQCD matrix elements is encoded in
\begin{equation} \label{eq:C2}
   C^{(2)}(x,\xi)=\frac{m_Q^2}{3} \langle v^2 \rangle_{V}\mathcal{M}_{\rho \sigma}^{(2)}I^{\rho \sigma}=\frac{\langle v^2 \rangle_{V}}{3 (2m_Q)} \frac{(x^2-7\xi^2)}{(x-\xi+i\epsilon)^2(x+\xi-i\epsilon)^2}
  \,.
\end{equation}
This is calculated from the second order term in Eq.(\ref{eq:amp_expansion}) (the gluon polarization already  projected onto the GPDs and factoring out the transverse metric tensor which contracts the external polarizations) and the projection tensor $I^{\rho \sigma}=-g^{\rho \sigma}+\frac{P_{V}^{\rho}P_{V}^{\sigma}}{4m_Q^2}$, coming from the sum over quarkonium polarization. Here $P_{V}$ is kept up to leading order in $k$ and therefore now satisfies $P_{V}^2=4m_Q^2$. 

\begin{figure}[htb]
    \centering
    \includegraphics[scale=1.4]{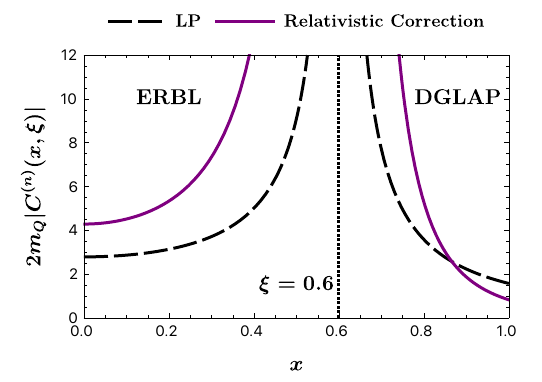}
    \caption{Plot of the absolute value of the matching coefficients in the dimensionless combination $2 m_Q |C^{(n)}|$ as a function of $x$. We show general photoproduction kinematics with both the ERBL and DGLAP regions for the choice $\xi=0.6$. The LP term $|C^{(0)}(x,\xi)|$ corresponds to the dashed (black) line while the relativistic correction $|C^{(2)}(x,\xi)|$ corresponds to the purple line (Only the purple line in the DGLAP region is negative). Here $\langle v^2 \rangle=0.22$ which is appropriate for $J/\psi$.}
    \label{fig:Match_Coeff}
\end{figure}

For the LP hard matching coefficient given in Eq.~(\ref{eq:C0}) the convolution integral with the GPD gives a finite result~\cite{Collins:1996fb}. This can be seen by decomposing the $x\pm\xi \mp i\epsilon$ factors into principal value and $\delta$-function contributions, and noting that both give finite results since the GPD is finite and continuous at the thresholds $x=\pm \xi$.
For the relativistic correction $C^{(2)}$ on the other hand these properties of the GPD do not suffice to make the convolution integrals finite due to the presence of the double poles at $x=\pm \xi$. This is shown in Fig.\ref{fig:Match_Coeff} where the matching coefficient for the $O(v^2)$ term diverges faster than the LP term at $x=\xi$.
Physically this is the transition point between the DGLAP and ERBL region where one of the gluons is soft.  These singularities at the endpoints of the two different kinematic regions are seen in other processes as well. The authors in \cite{Cui:2018jha} found similar endpoint divergences in exclusive photoproduction of HVM in the high energy limit when calculating higher twist corrections. Within the Soft-Collinear Effective Theory (SCET) framework \cite{Bauer:2000ew,Bauer:2000yr, Bauer:2001ct,Bauer:2001yt,Bauer:2002nz} these endpoint divergences have been found in computations of subleading power corrections for various processes \cite{Liu:2019oav,Beneke:2022obx,Luke:2022ops}. The cancellation of the divergences happens between different kinematic regions and can be handled by proper organization of components of the subleading power factorization theorem. For our process we expect an additional soft contribution that is active in the region $x\to \pm\xi$, but we leave the exploration and cancellation of these endpoint divergences for exclusive HVM photoproduction to future work. 
Based on the ubiquous presence of endpoint singularities in power corrections to exclusive hard scattering processes, we anticipate that they will also occur for corrections to the leading factorization that are not associated to the heavy quarks, which have not been considered here.  Thus, we expect that endpoint singularities will generically appear at next-to-leading power for processes involving either light quark or gluon GPDs.

On the other hand, for the near-threshold region, one must expand Eq.~(\ref{eq:C2}) about $|x|\ll |\xi|$, which gives
\begin{equation}
\label{eq:ExpAmp}
      \Amp[V]= \frac{8\sqrt{2}\pi g_{e} e_{Q} \alpha_s}{3} \left( \frac{\langle \mathcal{O}_{1}(^3S_1) \rangle_V}{M_V} \right)^{1/2} \left( \varepsilon_\gamma^\perp \cdot \varepsilon_V^{*\perp} \right) \left( 1-\frac{7\langle v^2 \rangle_{V}}{3} \right)  G^{(0)}(t,\xi) 
    .
\end{equation}
The small $x$ expansion circumvents the issue of the endpoint divergence, by expanding in a regime far from the dangerous pole.  
One can do a similar estimation as in (\ref{eq:onequarter}) of the contribution of the first moment of the relativistic correction to that of the higher moments. 
\begin{align}
    \frac{\sum_{n \ge 1}^{\infty} (6n+7)\int_{-1}^{1} dx \, \frac{x^{2n}}{\xi^{2n}}F_{g}^{\text{asym}}(x,\xi,t)}{7\int_{-1}^{1} dx \, F_{g}^{\text{asym}}(x,\xi,t)}=\frac{11}{14}\,.
\end{align}
Thus, unlike the LP terms, here the higher order moments contribute $\sim 78\%$, and only $22\%$ of the contribution of the relativistic contribution comes from the first moment. 
Thus a large number of higher moments terms are required to have a chance of obtaining convergence for the NLP contribution. In addition, as we will see in the next section, the zeroth moment of the relativistic correction is very sensitive to $\langle v^2\rangle $ and can lead to large cancellation of the LP contribution.
We will contrast this with the use of the full GPDs in the next section.

\section{Photoproduction cross-section}
\label{sec:Cross-section}

While we obtained the amplitude in the Ji frame, for easier comparison with previous work we calculate the cross-section in the proton-photon center-of-mass frame. The two frames are related by a rotation and a boost. Under the latter the collinear expansion is invariant while the former introduces a correction of the $O(t/W^2)$. This again is a kinematic power correction and thus is neglected in our calculation.
The differential cross-section with the relativistic NLP corrections is given by
\begin{align} \label{eq:GPDcross}
    \frac{d\sigma}{dt} &= \frac{1}{16\pi(W^2-M_N^2)^2}\frac{1}{2} \sum_{\text{polarization}}|\Amp|^2 \nonumber \\
    &=\frac{32\pi^2\alpha_{\text{EM}}e^2_{Q}\alpha_s^2}{9(W^2-M_N^2)^2}\left( \frac{M_V \langle \mathcal{O}_{1} \rangle_V}{4m_Q^2 (1+\langle v^2 \rangle_V)} \right)  \frac{1}{2} \sum_{\text{polarization}} \left| G(\xi,t) \right|^2
  ,
\end{align}
where $G(\xi,t)$ was defined in Eq.~(\ref{eq:RelCorrection}) and includes the full convolution with $C^{(0)}$ and $C^{(2)}$. If we keep only the LP term then we replace $G(\xi,t) \to G_{\rm LP}(\xi,t)$ and set $\langle v^2\rangle_V=0$.  Here the sum and averaging is over the polarization of the initial and final proton state.

After using the leading moment expansion  in Eq.(\ref{eq:ExpAmp}) the cross section simplifies to
\begin{align} \label{eq:moment_cross}
    \frac{d\sigma}{dt} &=\frac{32\pi^2\alpha_{\text{EM}}e^2_{Q}\alpha_s^2}{9(W^2-M_N^2)^2}\left( \frac{M_V \langle \mathcal{O}_{1} \rangle_V}{4m_Q^2 (1+\langle v^2 \rangle_V)} \right)\left( 1-\frac{7\langle v^2 \rangle_{V}}{3} \right)^2 \frac{1}{2} \sum_{\text{polarization}} \left| G^{(0)}(\xi,t) \right|^2
  ,
\end{align}
where $G^{(0)}(\xi,t)$ is given in Eq.~(\ref{eq:int_gpd}).
 Using the parametrization of the gluon GPD correlator in Eq.(\ref{eq:gluonGPD}), the sum/average of the squared moment of GPD has the form
\begin{align}
    \frac{1}{2} \sum_{\text{polarization}} \left| G^{(0)}(\xi,t) \right|^2 &= \frac{1}{m_Q^2\xi ^4} \Bigg[ \left(1-\xi ^2\right) (E_2(t,\xi )+H_2(t,\xi ))^2 \nonumber \\
    & -2 E_2(t,\xi) (E_2(t,\xi )+H_2(t,\xi ))+\left(1-\frac{t}{4 M_N^2}\right) E_2(t,\xi )^2 \Bigg] \,.
\end{align}

\begin{figure}[!ht]
    \centering
    \includegraphics[scale=1.2]{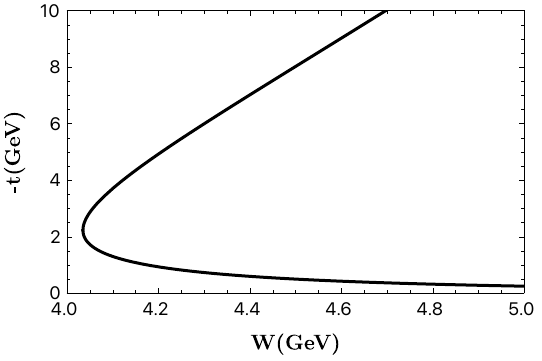}
    \caption{Kinematically allowed region in $(W,-t)$ plane for $J/\psi$}
    \label{fig:tWplot}
\end{figure}

To compare with GlueX data, we take the physical constants similar to the ones used in \cite{Guo:2023pqw}, with the proton mass $M_N=0.938$ GeV, $J/\psi$ mass $M_{J/\psi}=3.097$ GeV and $\alpha_s(\mu=2 \, \text{GeV})=0.3$. For the relativistic correction an additional parameter to consider is the charm quark mass which we take to be the pole mass $m_c=1.4 \, \text{GeV}$ . This gives $\langle v^2 \rangle_{J/\psi}=0.22$ from the Gremm-Kapustin relation. The other inputs for the cross-section are the nonperturbative matrix elements. The LDME $\langle \mathcal{O}_{1}(^3S_1) \rangle_{J/\psi}$ can be estimated from the leptonic decay rate of $J/\psi$ \cite{Eichten:2019hbb}
\begin{equation}
    \langle \mathcal{O}_{1}(^3S_1) \rangle_{J/\psi}=6 |\psi_{1S}(0)|^2=\frac{6.5712}{4\pi} \, \text{GeV}^3 \,.
\end{equation}
Finally, when using the moment expansion of the GPDs in Eq.~(\ref{eq:zeromoment}), the required gravitational form factors are taken from lattice calculations \cite{Pefkou:2021fni} with the $t$ dependence in the tripole expansion being
\begin{align}
    A_g(t) &=\frac{A_g(0)}{\left(1-\frac{t}{m_A}\right)^3}, 
 & 
    B_g(t) &=\frac{B_g(0)}{\left(1-\frac{t}{m_B}\right)^3}, 
 & 
    C_g(t) &=\frac{C_g(0)}{\left(1-\frac{t}{m_C}\right)^3}.
\end{align}
The values for the parameters in the above expression are: $A_g(0)=0.429,m_A=1.641\, \text{GeV}$, $B_g(0)=0.097, m_B=4.0 \, \text{GeV}$ and $C_g(0)=-1.93, m_C=1.07 \, \text{GeV}$. For the cross-section, we integrate between $-t_{-} <-t < -t_{+}$ as given in Eq. (\ref{eq:trange}) which corresponds to the bounded region in Fig.~\ref{fig:tWplot} for charmonium. In the threshold region, $\xi \rightarrow 1$, the upper and lower bounds coincide. Thus while we show the integrated cross-section near the threshold region they basically represent the differential cross-section.

\begin{figure}[!ht]
    \centering
    \includegraphics[scale=0.4]{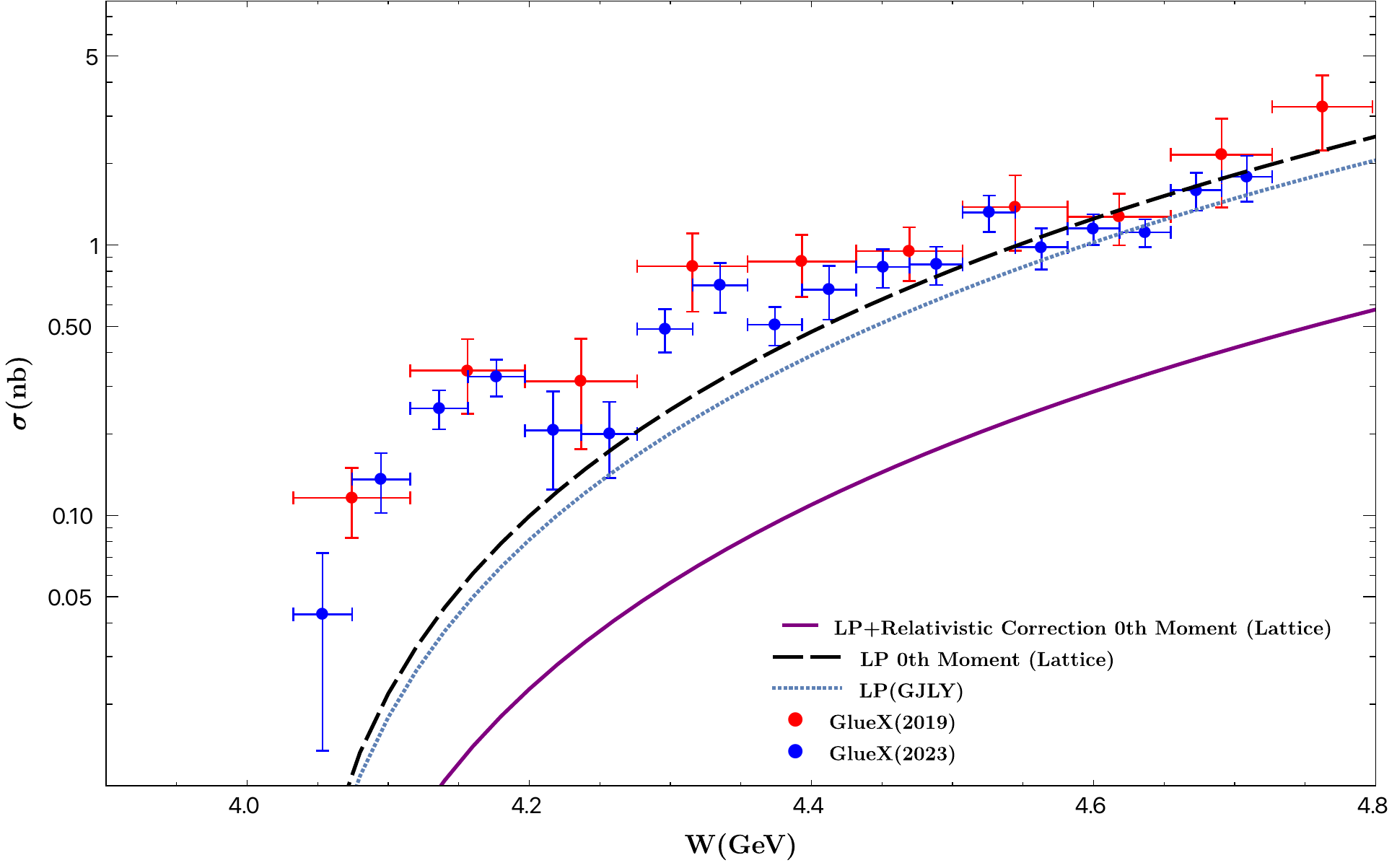}
    \caption{Log plot of the cross-section for $J/\psi$ production, where the threshold region starts from the smallest values of $W$ shown. The leading power (LP) and the relativistic correction to LP obtained from this work are shown in thick (purple) and dashed (black) lines respectively. The dotted line corresponds to LP result from \cite{Guo:2023pqw}.
    }
    \label{fig:Jpsi_moment}
\end{figure}

Using the GPD moment expansion as in Eq.~(\ref{eq:moment_cross}),
our results for the inclusion of relativistic corrections in the $J/\psi$ cross-section are shown in Fig.~\ref{fig:Jpsi_moment}. 
The first thing to note is that there is a small difference between the LP results obtained in this work (dashed black curve) and that of \cite{Guo:2023pqw} (dotted blue curve). This difference comes from the mass which was used in the hard matching coefficient. While here we used $2m_c$, the earlier work instead used  $M_{J/\psi}$. These coincide with each other only when the relative velocity between the heavy quark-antiquark pair is neglected, that is at LP. However, when adding relativistic effects these differ and for proper inclusion of the results at these different orders one should use the charm quark mass in the hard matching coefficient.
Secondly for a charm quark mass of $m_c=1.4$ GeV, one can see that there is a large cancellation between LP and NLP contributions, which reduces the cross section by about a factor of $5$. (This cancellation is even larger for $m_c=1.3\,{\rm GeV}$.)  This very large reduction is associated to the poor approximation of using only the zeroth moment of the GPD.

\begin{figure}[!ht]
    \centering
    \includegraphics[scale=0.4]{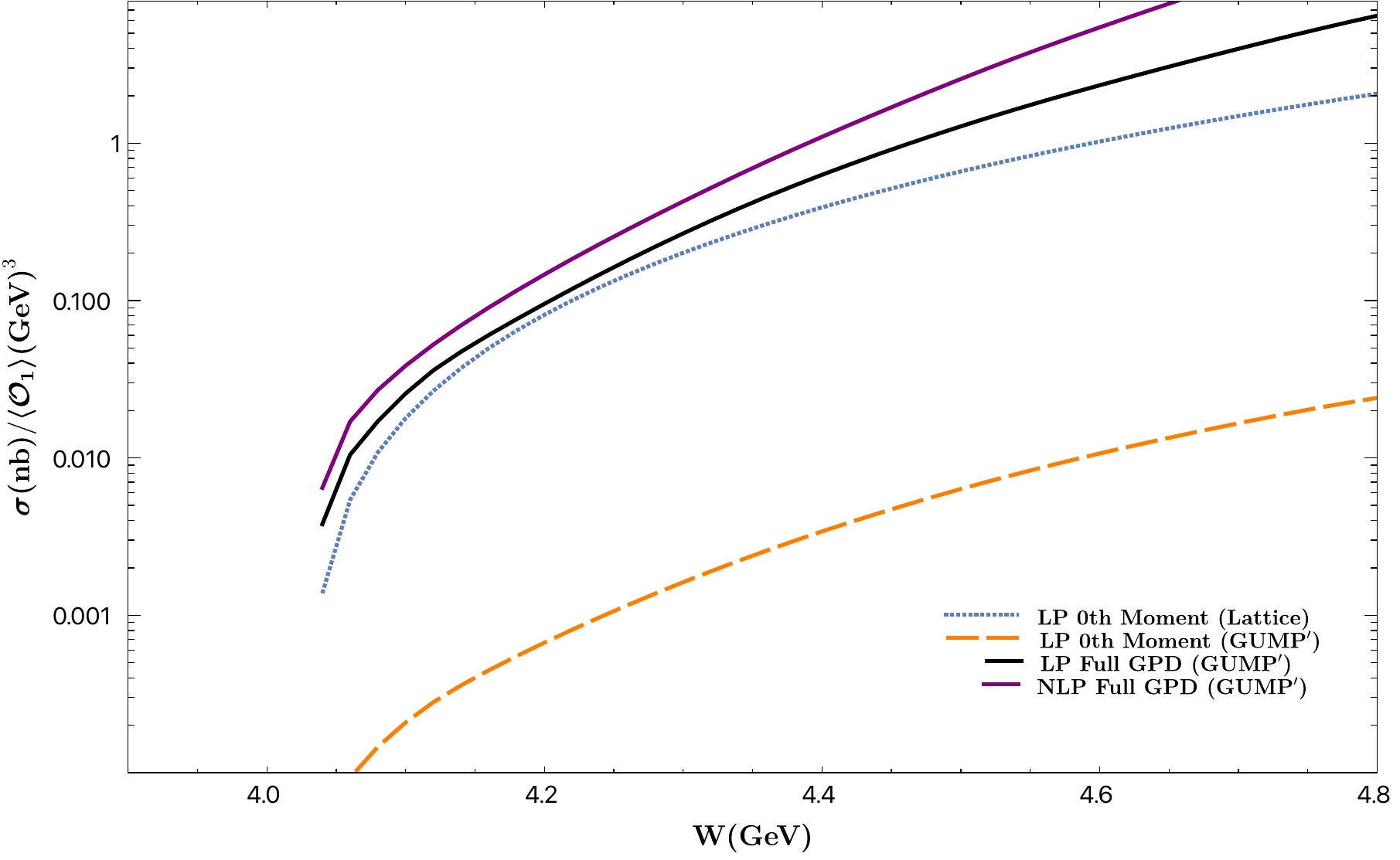}
    \caption{Log plot of the cross-section for $J/\psi$ production divided by the NRQCD LDME for the LP moments from lattice and GUMP$^\prime$ and full GPD dependent functions for LP and NLP contributions.}
    \label{fig:Scaled_Cross}
\end{figure}

To go beyond the moment approximation, we can consider model based results for the full GPD functions, whose parameters are to be constrained by experimental data. As a test of this approach, 
we  use a modified version of the GUMP \cite{Guo:2025muf} which we will call as GUMP$^\prime$ henceforth. In this model the gluon GPDs $H_g(x,\xi,t)$ and $E_g(x,\xi,t)$ extracted from GUMP are normalized by tunable parameters $N_H$ and $N_E$. DVMP data for $J/\psi$ was not used to extract the GUMP gluon GPDs and thus moments of gluon GPDs do not match the ones obtained from the tripole expansion of lattice in shape or magnitude.
The addition of the parameters $N_H$ and $N_E$ provides us with a handle to compensate for this, and get the theoretical cross-section at LP close to data.
The convergence of moments in the GUMP$^\prime$ model is poor even at LP, which can be seen by comparing the dashed-yellow and solid-black curves in Fig.~\ref{fig:Scaled_Cross}.
Here the effects of the NRQCD LDME is removed by dividing it out of the cross-section.

Using GUMP$^\prime$ one can now assess the impact of the NLP relativistic correction with respect to the LP result. Here the cross-section is obtained from Eq.~(\ref{eq:GPDcross}) without expanding the matching coefficients.
With the GUMP$^\prime$ model and full convolution with the hard coefficients, one can see from Fig.~\ref{fig:Scaled_Cross} that the NLP corrections are still significant, but do not have as dramatic an impact as was seen in the moment based approach.

\begin{figure}[!ht]
    \centering
    \includegraphics[scale=0.4]{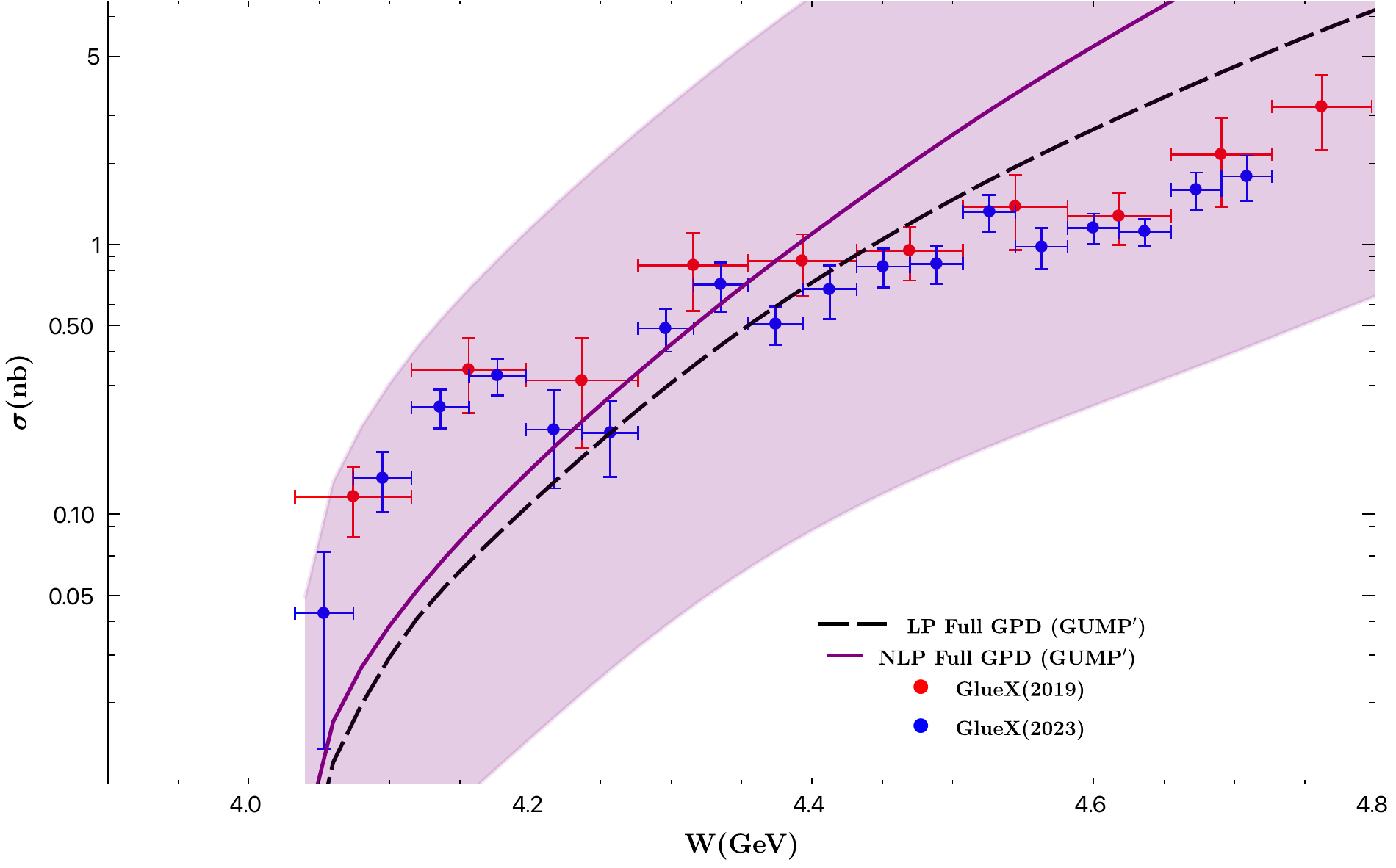}
    \caption{Log plot of the cross-section for $J/\psi$ production. The leading power (LP) and the relativistic correction to LP obtained from this work are shown in thick (purple) and dashed (black) lines respectively.
    }
    \label{fig:Jpsi}
\end{figure}

\begin{figure}[!ht]
    \centering
    \includegraphics[scale=0.4]{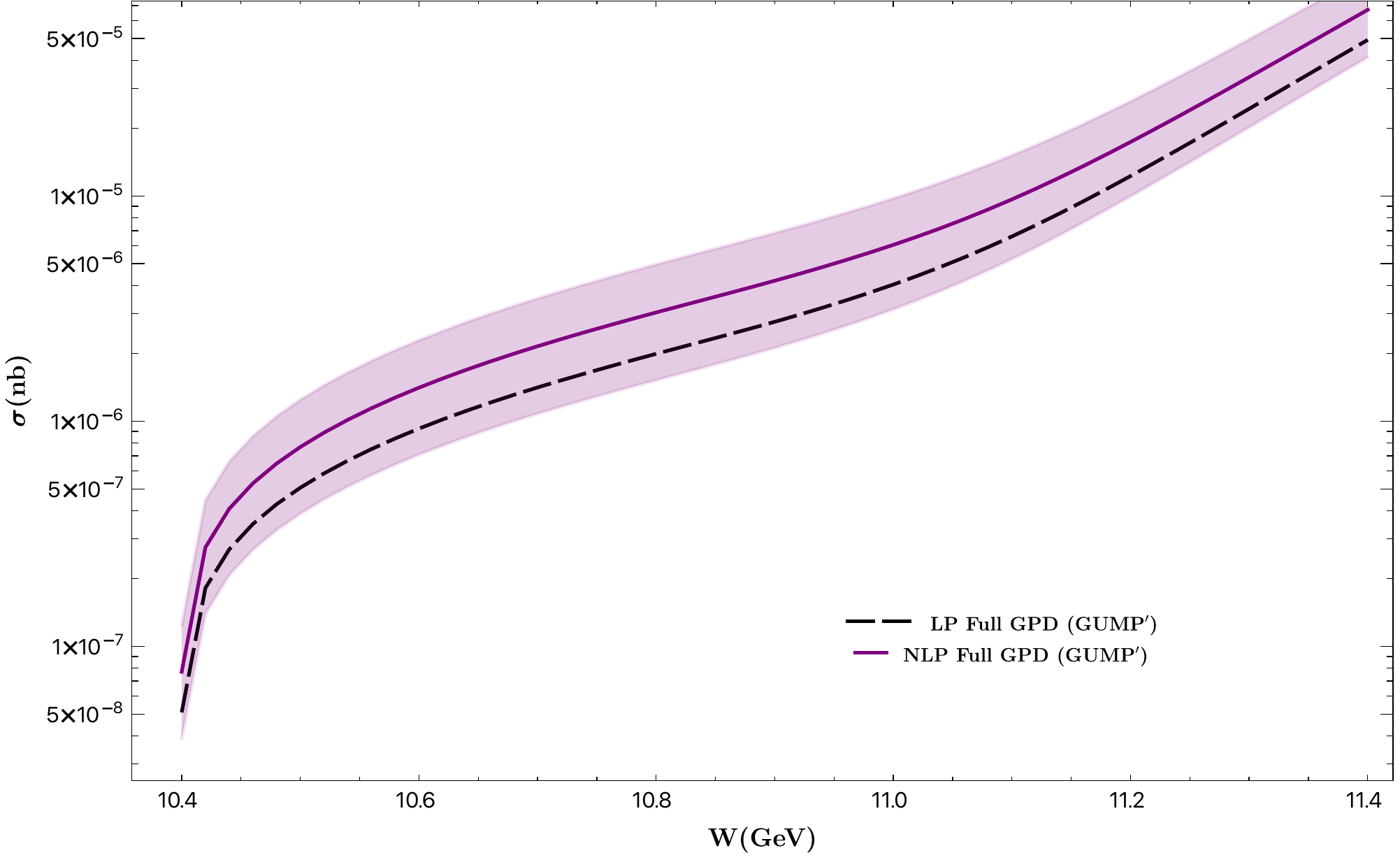}
    \caption{Log plot of the cross-section for $\Upsilon$ production. Mass parameters used for this plot are $M_\Upsilon=9.46$ GeV and $m_b=4.78$ GeV which gives $\langle v^2 \rangle_{\Upsilon}=0.013$. 
    The purple shaded region gives an estimates for the uncertainty from missing power corrections that are of the same size as those computed here.
   Here the LDME estimated from leptonic decay is taken to be $\langle \mathcal{O}_{1}(^3S_1) \rangle_{\Upsilon}=\frac{35.1528}{4\pi} \, \text{GeV}^3$. }
    \label{fig:Upsilon}
\end{figure}

The results for the cross-section with this approach are compared to data for $J/\Psi$ in Fig~\ref{fig:Jpsi}. To obtain the LP cross-section we have taken $N_H=0.1$ and $N_E=6.0$.
We also use our results to make a prediction for the Upsilon cross-section in Fig.~\ref{fig:Upsilon}, and give the corresponding parameter values in the figure caption.
The cross-section at NLP is obtained by integrating the GPDs against hard functions with double poles at $x=\pm \xi$. 
Near $x=\pm\xi$ collinear partons become soft, and so it is sensible to cutoff this region from the convolution integral.
Thus we integrate $x$ from $0$ to $(1-\lambda) \xi$ and $(1+\lambda)\xi$ to 1, where $\lambda$ is a dimensionless cutoff. For the central solid-purple curve we have chosen $\lambda=0.03$.
In our predictions the two dominant sources are (i) the uncertainty in the value of the charm quark mass $\Delta\sigma_{\text{mass}}$ which also leads to uncertainty in the matrix element $\langle v^2 \rangle$,  and (ii) choice for the NLP cutoff parameter $\lambda$. The error $\Delta\sigma_{\text{mass}}$ is estimated  by taking $m_c=1.4 \pm 0.1 \, \text{GeV}$, since our analysis is only at leading order in $\alpha_s$.
The error $\Delta\sigma_{\lambda}$ is due to variation of the cross-section when taking $\lambda=0.03^{+.06}_{-.02}$. 
The final uncertainty band is then $\Delta\sigma =\sqrt{(\Delta\sigma_{\lambda})^2+(\Delta\sigma_{\text{mass}})^2}$, and is quite large. 
Here the dominant effect comes from the charm mass and corresponding 
variation of $\langle v^2 \rangle_{J/\psi}$. This uncertainty is reduced in the case of $\Upsilon$ production with $m_b = 4.78 \pm 0.08 \, {\rm GeV}$ as can be seen from Fig.~\ref{fig:Upsilon}, because $v^2$ is smaller in the bottom system.

\section{Conclusion}
\label{sec:Summary}

We have calculated the relativistic correction to exclusive HVM production in the near-threshold limit in the GPD+NRQCD framework. In general these relativistic corrections are found to be sizable for $J/\psi$ production in other processes since $v^2 \sim \alpha_s(2m_c)$ is relatively large. Thus relativistic corrections to $J/\psi$ production are as important as NLO perturbative contributions. We find this to be true in exclusive photoproduction as well. 

The GPD moment expansion, while expected to work near the threshold region, fails due to the finite mass of the heavy quark, leading to $1-\xi\sim M_N/M_V$. 
At NLP the 
corrections from the higher moments are very large, so the result utilizing the full GPD dependence is more reliable and differs a lot compared to the moment result. Even using the full GPD, the NLP relativistic corrections change the cross sections by about a factor of two. The results are also sensitive to the charm quark mass and therefore inclusion of higher order corrections in $\alpha_s$ at both LP and NLP will be required for better extraction of the GPD.
We also make predictions for $\Upsilon$ production where relativistic corrections are smaller. 
This provides an important method to test and constrain gluon GPDs at the future EIC collider.

In terms of future work there are several issues to address. The first is to investigate other sources of power corrections for the complete picture at next-to-leading power. The second is to resolve the issue of endpoint divergence at subleading power in the context of photo/electroproduction away from near-threshold kinematics. Finally, $\alpha_s$ corrections at NLP should be computed. We intend to follow up on these topics in future work.

\acknowledgments

We thank Maxim Nefedov for pointing out an overall sign error in our expression for the NLP matching coefficient in a previous draft.  We acknowledge support from the U.S. Department of Energy, Office of Science, Office of Nuclear Physics under grant contract numbers DE-FG02-04ER41338, DE-FG02-05ER41367 and DE-SC0011090. This work was also supported in part by DOE Quark-Gluon Tomography Topical Collaboration with award number DE-SC0023646, including direct support for 
SB and JR. 
IS is also supported in part by Simons Foundation through the Investigator grant 327942, and 
thanks the Delta ITP consortium, a program of the Netherlands Organisation for
Scientific Research (NWO) that is funded by the Dutch Ministry of Education, Culture and Science (OCW) for support, and the Nikhef and DESY theory groups for hospitality during the completion of this work.
The Feynman diagrams in this work were evaluated using FeynCalc \cite{Shtabovenko:2020gxv,Shtabovenko:2016sxi,Mertig:1990an}.

Finally we would like to dedicate this work to the memory of Thomas Mehen (1970-2024) and Fanyi Zhao (1996-2024), both of whom tragically passed away before the publication of this work.

\bibliographystyle{JHEP}
\bibliography{biblio.bib}

\end{document}